\documentclass[12pt]{article}
\usepackage{graphicx}
\usepackage{cite}
\usepackage{amsmath}
\usepackage{hyperref}
\usepackage{graphicx}
\usepackage{amssymb}
\usepackage{amsmath}
\usepackage{epsfig}

\begin{document}


\newcommand{\be}{\begin{equation}}
\newcommand{\beq}{\begin{equation}}
\newcommand{\eeq}{\end{equation}}
\newcommand{\ee}{\end{equation}}

\newcommand{\refeq}[1]{Eq.\ref{eq:#1}}
\newcommand{\refig}[1]{Fig.\ref{fig:#1}}
\newcommand{\refsec}[1]{Sec.\ref{sec:#1}}

\newcommand{\beqn}{\begin{eqnarray}}
\newcommand{\eeqn}{\end{eqnarray}}
\newcommand{\bea}{\begin{eqnarray}}
\newcommand{\ena}{\end{eqnarray}}
\newcommand{\ra}{\rightarrow}
\newcommand{\susy}{{{\cal SUSY}$\;$}}
\newcommand{\su}{$ SU(2) \times U(1)\,$}

\newcommand{\gag}{$\gamma \gamma$ }
\newcommand{\gagt}{\gamma \gamma }
\newcommand{\gam}{\gamma \gamma }
\def\W{{\mbox{\boldmath $W$}}}
\def\B{{\mbox{\boldmath $B$}}}
\def\V{{\mbox{\boldmath $V$}}}
\newcommand{\np}{Nucl.\,Phys.\,}
\newcommand{\pl}{Phys.\,Lett.\,}
\newcommand{\pr}{Phys.\,Rev.\,}
\newcommand{\prl}{Phys.\,Rev.\,Lett.\,}
\newcommand{\prep}{Phys.\,Rep.\,}
\newcommand{\zp}{Z.\,Phys.\,}
\newcommand{\sovjnp}{{\em Sov.\ J.\ Nucl.\ Phys.\ }}
\newcommand{\nuclinst}{{\em Nucl.\ Instrum.\ Meth.\ }}
\newcommand{\annp}{{\em Ann.\ Phys.\ }}
\newcommand{\intjmp}{{\em Int.\ J.\ of Mod.\  Phys.\ }}

\newcommand{\eps}{\epsilon}
\newcommand{\mw}{M_{W}}
\newcommand{\mww}{M_{W}^{2}}
\newcommand{\mwmw}{M_{W}^{2}}
\newcommand{\mhmh}{M_{H}^2}
\newcommand{\mz}{M_{Z}}
\newcommand{\mzz}{M_{Z}^{2}}

\newcommand{\cw}{c_W}
\newcommand{\sw}{s_W}
\newcommand{\tw}{\tan\theta_W}
\def\tww{\tan^2\theta_W}
\def\stw{s_{2w}}

\newcommand{\smw}{s_M^2}
\newcommand{\cmw}{c_M^2}
\newcommand{\seff}{s_{{\rm eff}}^2}
\newcommand{\ceff}{c_{{\rm eff}}^2}
\newcommand{\seffl}{s_{{\rm eff\;,l}}^{2}}
\newcommand{\sww}{s_W^2}
\newcommand{\cww}{c_W^2}
\newcommand{\swo}{s_W}
\newcommand{\cwo}{c_W}

\newcommand{\epm}{$e^{+} e^{-}\;$}
\newcommand{\epemt}{$e^{+} e^{-}\;$}
\newcommand{\epem}{e^{+} e^{-}\;}
\newcommand{\ememt}{$e^{-} e^{-}\;$}
\newcommand{\emem}{e^{-} e^{-}\;}

\newcommand{\ord}{{\cal O}}

\newcommand{\lra}{\leftrightarrow}
\newcommand{\tr}{{\rm Tr}}
\def\ls1{{\not l}_1}
\newcommand{\cms}{centre-of-mass\hspace*{.1cm}}


\newcommand{\dkg}{\Delta \kappa_{\gamma}}
\newcommand{\dkz}{\Delta \kappa_{Z}}
\newcommand{\dz}{\delta_{Z}}
\newcommand{\dgz}{\Delta g^{1}_{Z}}
\newcommand{\dgzt}{$\Delta g^{1}_{Z}\;$}
\newcommand{\la}{\lambda}
\newcommand{\lag}{\lambda_{\gamma}}
\newcommand{\lambdae}{\lambda_{e}}
\newcommand{\laz}{\lambda_{Z}}
\newcommand{\lnl}{L_{9L}}
\newcommand{\lnr}{L_{9R}}
\newcommand{\lt}{L_{10}}
\newcommand{\lu}{L_{1}}
\newcommand{\ld}{L_{2}}
\newcommand{\eeww}{e^{+} e^{-} \ra W^+ W^- \;}
\newcommand{\eewwt}{$e^{+} e^{-} \ra W^+ W^- \;$}
\newcommand{\epemww}{e^{+} e^{-} \ra W^+ W^- }
\newcommand{\epemwwt}{$e^{+} e^{-} \ra W^+ W^- \;$}
\newcommand{\eennhht}{$e^{+} e^{-} \ra \nu_e \bar \nu_e HH\;$}
\newcommand{\eennhh}{e^{+} e^{-} \ra \nu_e \bar \nu_e HH\;}
\newcommand{\eennht}{$e^{+} e^{-} \ra \nu_e \bar \nu_e H\;$}
\newcommand{\eennh}{e^{+} e^{-} \ra \nu_e \bar \nu_e H\;}
\newcommand{\ppwg}{p p \ra W \gamma}
\newcommand{\wwhh}{W^+ W^- \ra HH\;}
\newcommand{\wwhht}{$W^+ W^- \ra HH\;$}
\newcommand{\ppwz}{pp \ra W Z}
\newcommand{\ppwgt}{$p p \ra W \gamma \;$}
\newcommand{\ppwzt}{$pp \ra W Z \;$}
\newcommand{\gamgamt}{$\gamma \gamma \;$}
\newcommand{\gamgam}{\gamma \gamma \;}
\newcommand{\egamt}{$e \gamma \;$}
\newcommand{\egam}{e \gamma \;}
\newcommand{\gamgamwwt}{$\gamma \gamma \ra W^+ W^- \;$}
\newcommand{\gamgamwwht}{$\gamma \gamma \ra W^+ W^- H \;$}
\newcommand{\gamgamwwh}{\gamma \gamma \ra W^+ W^- H \;}
\newcommand{\gamgamwwhht}{$\gamma \gamma \ra W^+ W^- H H\;$}
\newcommand{\gamgamwwhh}{\gamma \gamma \ra W^+ W^- H H\;}
\newcommand{\ggww}{\gamma \gamma \ra W^+ W^-}
\newcommand{\ggwwt}{$\gamma \gamma \ra W^+ W^- \;$}
\newcommand{\ggwwht}{$\gamma \gamma \ra W^+ W^- H \;$}
\newcommand{\ggwwh}{\gamma \gamma \ra W^+ W^- H \;}
\newcommand{\ggwwhht}{$\gamma \gamma \ra W^+ W^- H H\;$}
\newcommand{\ggwwhh}{\gamma \gamma \ra W^+ W^- H H\;}
\newcommand{\ggwwz}{\gamma \gamma \ra W^+ W^- Z\;}
\newcommand{\ggwwzt}{$\gamma \gamma \ra W^+ W^- Z\;$}

\newcommand{\veps}{\varepsilon}

\newcommand{\ptu}{p_{1\bot}}
\newcommand{\vecptu}{\vec{p}_{1\bot}}
\newcommand{\ptd}{p_{2\bot}}
\newcommand{\vecptd}{\vec{p}_{2\bot}}
\newcommand{\ie}{{\em i.e.}}
\newcommand{\cm}{{{\cal M}}}
\newcommand{\cl}{{{\cal L}}}
\newcommand{\cd}{{{\cal D}}}
\newcommand{\cv}{{{\cal V}}}
\def\slashc{c\kern -.400em {/}}
\def\slashp{p\kern -.400em {/}}
\def\slashq{q\kern -.450em {/}}
\def\slashL{L\kern -.450em {/}}
\def\slashcl{\cl\kern -.600em {/}}
\def\slashr{r\kern -.450em {/}}
\def\slashk{k\kern -.500em {/}}
\def\Ww{{\mbox{\boldmath $W$}}}
\def\B{{\mbox{\boldmath $B$}}}
\def\noi{\noindent}
\def\nn{\noindent}
\def\sm{${\cal{S}} {\cal{M}}\;$}
\def\smn{${\cal{S}} {\cal{M}}$}
\def\smp{${\cal{S}} {\cal{M}}$}
\def\nph{${\cal{N}} {\cal{P}}\;$}
\def\sb{$ {\cal{S}}  {\cal{B}}\;$}
\def\ssb{${\cal{S}} {\cal{S}}  {\cal{B}}\;$}
\def\ssbe{{\cal{S}} {\cal{S}}  {\cal{B}}}
\def\cviol{${\cal{C}}\;$}
\def\pviol{${\cal{P}}\;$}
\def\cpviol{${\cal{C}} {\cal{P}}\;$}

\newcommand{\lgg}{\lambda_1\lambda_2}
\newcommand{\lww}{\lambda_3\lambda_4}
\newcommand{\ppin}{ P^+_{12}}
\newcommand{\pmin}{ P^-_{12}}
\newcommand{\ppout}{ P^+_{34}}
\newcommand{\pmout}{ P^-_{34}}
\newcommand{\sinsq}{\sin^2\theta}
\newcommand{\cossq}{\cos^2\theta}
\newcommand{\yt}{y_\theta}
\newcommand{\hppll}{++;00}
\newcommand{\hpmll}{+-;00}
\newcommand{\hpplt}{++;\lambda_30}
\newcommand{\hpmlt}{+-;\lambda_30}
\newcommand{\hpptt}{++;\lambda_3\lambda_4}
\newcommand{\hpmtt}{+-;\lambda_3\lambda_4}
\newcommand{\dk}{\Delta\kappa}
\newcommand{\klam}{\Delta\kappa \lambda_\gamma }
\newcommand{\kac}{\Delta\kappa^2 }
\newcommand{\lac}{\lambda_\gamma^2 }
\def\gamgamtzz{$\gamma \gamma \ra ZZ \;$}
\def\gamgamtww{$\gamma \gamma \ra W^+ W^-\;$}
\def\gamgamtwwe{\gamma \gamma \ra W^+ W^-}

\def\intfd{ \int \frac{d^4 r}{(2\pi)^4} }
\def\intnd{ \int \frac{d^n r}{(2\pi)^n} }
\def\intnmu{ \mu^{4-n} \int \frac{d^n r}{(2\pi)^n} }
\newcommand{\Dkm}{[(r+k)^2-m_2^2]}
\newcommand{\Dkom}{[(r+k_1)^2-m_2^2]}
\newcommand{\Dkotm}{[(r+k_1+k_2)^2-m_3^2]}
\def\piggt{$\Pi_{\gamma \gamma}\;$}
\def\pigg{\Pi_{\gamma \gamma}}
\newcommand{\mn}{{\mu \nu}}
\newcommand{\mzb}{M_{Z,0}}
\newcommand{\mzbs}{M_{Z,0}^2}
\newcommand{\mwb}{M_{W,0}}
\newcommand{\mwbs}{M_{W,0}^2}
\newcommand{\dgg}{\frac{\delta g^2}{g^2}}
\newcommand{\dee}{\frac{\delta e^2}{e^2}}
\newcommand{\dss}{\frac{\delta s^2}{s^2}}
\newcommand{\dmw}{\frac{\delta \mww}{\mww}}
\newcommand{\dmz}{\frac{\delta \mzz}{\mzz}}
\def\pigz{\Pi_{\gamma Z}}
\def\pizz{\Pi_{Z Z}}
\def\piww{\Pi_{WW}}
\def\pioo{\Pi_{11}}
\def\pitt{\Pi_{33}}
\def\pitq{\Pi_{3Q}}
\def\piqq{\Pi_{QQ}}
\def\delr{\Delta r}
\def\calm{{\cal {M}}}
\def\gww{G_{WW}}
\def\gzz{G_{ZZ}}
\def\goo{G_{11}}
\def\gtt{G_{33}}
\def\szz{s_Z^2}
\def\estk{e_\star^2(k^2)}
\def\sstk{s_\star^2(k^2)}
\def\cstk{c_\star^2(k^2)}
\def\sstz{s_\star^2(\mzz)}
\def\mzst{{M_Z^{\star}}(k^2)^2}
\def\mwst{{M_W^{\star}}(k^2)^2}
\def\epo{\varepsilon_1}
\def\epd{\varepsilon_2}
\def\ept{\varepsilon_3}
\def\dro{\Delta \rho}
\def\gmu{G_\mu}
\def\alpz{\alpha_Z}
\def\danpmz{\Delta\alpha_{{\rm NP}}(\mzz)}
\def\danpk{\Delta\alpha_{{\rm NP}}(k^2)}
\def\calt{{\cal {T}}}
\def\piggh{\pigg^h(s)}
\def\cuv{C_{UV}}
\def\pilr{G_{LR}}
\def\pill{G_{LL}}
\def\dak{\Delta \alpha(k^2)}
\def\damz{\Delta \alpha(\mzz)}
\def\dahmz{\Delta \alpha^{(5)}_{{\rm had}}(\mzz)}
\def\sth{s_{\theta}^2}
\def\cth{c_{\theta}^2}
\newcommand{\siki}[1]{Eq.\ref{eq:#1}}
\newcommand{\zu}[1]{Fig.\ref{fig:#1}}
\newcommand{\setu}[1]{Sec.\ref{sec:#1}}
\newcommand{\anlg}{\tilde\alpha}
\newcommand{\bnlg}{\tilde\beta}
\newcommand{\dnlg}{\tilde\delta}
\newcommand{\enlg}{\tilde\varepsilon}
\newcommand{\knlg}{\tilde\kappa}
\newcommand{\xiw}{\xi_W}
\newcommand{\xiz}{\xi_Z}
\newcommand{\dbr}{\delta_B}
\newcommand{\bothd}{{ \leftrightarrow \atop{\partial^{\mu}} } }

\newcommand{\BARE}[1]{\underline{#1}}
\newcommand{\ZF}[1]{\sqrt{Z}_{#1}}
\newcommand{\ZFT}[1]{\tilde{Z}_{#1}}
\newcommand{\ZH}[1]{\delta Z_{#1}^{1/2}}
\newcommand{\ZHb}[1]{\delta Z_{#1}^{1/2\,*}}
\newcommand{\DM}[1]{\delta M^2_{#1}}
\newcommand{\DMS}[1]{\delta M_{#1}}
\newcommand{\Dm}[1]{\delta m_{#1}}
\newcommand{\tree}[1]{\langle {#1}\rangle}

\newcommand{\Cuv}{C_{UV}}
\newcommand{\logw}{\log M_W^2}
\newcommand{\logz}{\log M_Z^2}
\newcommand{\logh}{\log M_H^2}
\newcommand{\swt}{s_W^2}
\newcommand{\cwt}{c_W^2}
\newcommand{\swf}{s_W^4}
\newcommand{\cwf}{c_W^4}
\newcommand{\MWt}{M_W^2}
\newcommand{\MZt}{M_Z^2}
\newcommand{\MHt}{M_H^2}

\newcommand{\VECsl}[1]{\not{#1}}

\newcommand{\Bphi}{\mbox{\boldmath$\phi$}}
\newcommand{\eetth}{$e^+ e^-\ra t \bar{t} H$}
\newcommand{\eettht}{$e^+ e^-\ra t \bar{t} H\;$}
\newcommand{\nnhet}{$\epem \ra \nu_e \bar{\nu}_e H \;$}
\newcommand{\nnhe}{$\epem \ra \nu_e \bar{\nu}_e H$}
\newcommand{\eezh}{$\epem \ra Z H$}
\newcommand{\eezht}{$\epem \ra Z H \;$}
\newcommand{\eezhh}{$\epem \ra Z  H H$}
\newcommand{\eezhht}{$\epem \ra Z H H\;$}
\newcommand{\eeeeht}{$\epem \ra e^+ e^- H \;$}
\newcommand{\eeeeh}{$\epem \ra e^+ e^-  H$}
\newcommand{\eenngt}{$\epem \ra e^+ e^- \gamma \;$}
\newcommand{\eenng}{$\epem \ra e^+ e^-  \gamma$}

\def\al{\alpha}
\def\bt{\beta}
\def\gm{\gamma}
\def\Gm{\Gamma}
\def\et{\eta}
\def\del{\delta}
\def\Del{\Delta}
\def\kp{\kappa}
\def\lm{\lambda}
\def\Lm{\Lambda}
\def\th{\theta}
\def\zt{\zeta}
\def\ro{\rho}
\def\sig{\sigma}
\def\Sig{\Sigma}
\def\eps{\epsilon}
\def\vare{\varepsilon}
\def\vphi{\varphi}
\def\om{\omega}
\def\Om{\Omega}
\def\bar{\overline}
\def\d{{\rm d}}
\def\pdf{\partial}
\def\Int{\int\nolimits}
\def\det{{\rm det}}
\def\non{\nonumber}
\def\eqn{\begin{equation}}
\def\eqne{\end{equation}}
\def\eqa{\begin{eqnarray}}
\def\eqae{\end{eqnarray}}
\def\ary{\begin{array}}
\def\arye{\end{array}}
\def\dsc{\begin{description}}
\def\dsce{\end{description}}
\def\itm{\begin{itemize}}
\def\itme{\end{itemize}}
\def\enu{\begin{enumerate}}
\def\enue{\end{enumerate}}
\def\ct{\begin{center}}
\def\cte{\end{center}}
\def\D{{\cal D}}
\def\bfD{{\bf D}}

\newcommand{\cha}{{tt CHANEL}}

\def\sinb{\sin\beta}
\def\cosb{\cos\beta}
\def\sinbb{s_ {2\beta}}
\def\cosbb{c_{2 \beta}}
\def\tgb{\tan \beta}
\def\tgbt{$\tan \beta\;\;$}
\def\tgbsq{\tan^2 \beta}
\def\tgbsqt{$\tan^2 \beta\;$}
\def\sinal{\sin\alpha}
\def\cosal{\cos\alpha}
\def\sb{s_\beta}
\def\cb{c_\beta}
\def\tb{t_\beta}
\def\ttb{t_{2 \beta}}
\def\sa{s_\alpha}
\def\ca{c_\alpha}
\def\ta{t_\alpha}
\def\stb{s_{2\beta}}
\def\ctb{c_{2\beta}}
\def\sbb{s_ {2\beta}}
\def\cbb{c_{2 \beta}}
\def\sta{s_{2\alpha}}
\def\cta{c_{2\alpha}}
\def\sbma{s_{\beta-\alpha}}
\def\cbma{c_{\beta-\alpha}}
\def\sbpa{s_{\beta+\alpha}}
\def\cbpa{c_{\beta+\alpha}}
\def\lone{\lambda_1}
\def\ltwo{\lambda_2}
\def\lthree{\lambda_3}
\def\lfour{\lambda_4}
\def\lfive{\lambda_5}
\def\lsix{\lambda_6}
\def\lseven{\lambda_7}
\def\stop{\tilde{t}}
\def\sto{\tilde{t}_1}
\def\stt{\tilde{t}_2}
\def\stl{\tilde{t}_L}
\def\str{\tilde{t}_R}
\def\msto{m_{\sto}}
\def\mstosq{m_{\sto}^2}
\def\mstt{m_{\stt}}
\def\msttsq{m_{\stt}^2}
\def\mt{m_t}
\def\mtsq{m_t^2}
\def\sint{\sin\theta_{\stop}}
\def\sintt{\sin 2\theta_{\stop}}
\def\cost{\cos\theta_{\stop}}
\def\sintsq{\sin^2\theta_{\stop}}
\def\costsq{\cos^2\theta_{\stop}}
\def\mqtt{\M_{\tilde{Q}_3}^2}
\def\mutt{\M_{\tilde{U}_{3R}}^2}
\def\sbottom{\tilde{b}}
\def\sbo{\tilde{b}_1}
\def\sbt{\tilde{b}_2}
\def\sbl{\tilde{b}_L}
\def\sbr{\tilde{b}_R}
\def\msbo{m_{\sbo}}
\def\msbosq{m_{\sbo}^2}
\def\msbt{m_{\sbt}}
\def\msbtsq{m_{\sbt}^2}
\def\mt{m_t}
\def\mtsq{m_t^2}
\def\selectron{\tilde{e}}
\def\seo{\tilde{e}_1}
\def\set{\tilde{e}_2}
\def\sel{\tilde{e}_L}
\def\ser{\tilde{e}_R}
\def\mseo{m_{\seo}}
\def\mseosq{m_{\seo}^2}
\def\mset{m_{\set}}
\def\msetsq{m_{\set}^2}
\def\msel{m_{\sel}}
\def\mser{m_{\ser}}
\def\me{m_e}
\def\mesq{m_e^2}
\def\snu{\tilde{\nu}}
\def\snue{\tilde{\nu_e}}
\def\set{\tilde{e}_2}
\def\snul{\tilde{\nu}_L}
\def\msnue{m_{\snue}}
\def\msnuesq{m_{\snue}^2}
\def\smuon{\tilde{\mu}}
\def\smul{\tilde{\mu}_L}
\def\smur{\tilde{\mu}_R}
\def\msmul{m_{\smul}}
\def\msmulsq{m_{\smul}^2}
\def\msmur{m_{\smur}}
\def\msmursq{m_{\smur}^2}
\def\stau{\tilde{\tau}}
\def\stauo{\tilde{\tau}_1}
\def\staut{\tilde{\tau}_2}
\def\staul{\tilde{\tau}_L}
\def\staur{\tilde{\tau}_R}
\def\mstauo{m_{\stauo}}
\def\mstauosq{m_{\stauo}^2}
\def\mstaut{m_{\staut}}
\def\mstautsq{m_{\staut}^2}
\def\mtau{m_\tau}
\def\mtausq{m_\tau^2}
\def\gluino{\tilde{g}}
\def\mgluino{m_{\tilde{g}}}
\def\mchi{m_\chi^+}
\def\neuto{\tilde{\chi}_1^0}
\def\mneuto{m_{\tilde{\chi}_1^0}}
\def\neutt{\tilde{\chi}_2^0}
\def\mneutt{m_{\tilde{\chi}_2^0}}
\def\neutth{\tilde{\chi}_3^0}
\def\mneutth{m_{\tilde{\chi}_3^0}}
\def\neutf{\tilde{\chi}_4^0}
\def\mneutf{m_{\tilde{\chi}_4^0}}
\def\chargop{\tilde{\chi}_1^+}
\def\mchargo{m_{\tilde{\chi}_1^+}}
\def\chargtp{\tilde{\chi}_2^+}
\def\mchargt{m_{\tilde{\chi}_2^+}}
\def\chargom{\tilde{\chi}_1^-}
\def\chargtm{\tilde{\chi}_2^-}
\def\bino{\tilde{b}}
\def\wino{\tilde{w}}
\def\photino{\tilde{\gamma}}
\def\zino{tilde{z}}
\def\sdowno{\tilde{d}_1}
\def\sdownt{\tilde{d}_2}
\def\sdownl{\tilde{d}_L}
\def\sdownr{\tilde{d}_R}
\def\supo{\tilde{u}_1}
\def\supt{\tilde{u}_2}
\def\supl{\tilde{u}_L}
\def\supr{\tilde{u}_R}
\def\mh{m_h}
\def\mht{m_h^2}
\def\MH{M_H}
\def\MHt{M_H^2}
\def\MA{M_A}
\def\MAt{M_A^2}
\def\MHp{M_H^+}
\def\MHm{M_H^-}
\def\mqt{\M_{\tilde{Q}_3}}
\def\mut{\M_{\tilde{U}_{3R}}}
\def\mqtz{\M_{\tilde{Q}_3(0)}}
\def\mutz{\M_{\tilde{U}_{3R}(0)}}
\def\mqtzt{\M_{\tilde{Q}_3^2(0)}}
\def\mutzt{\M_{\tilde{U}_{3R}^2(0)}}

\def\mhf{M_{1/2}}

\begin{titlepage}

\begin{center}

\vspace*{5cm}

{\Large {\bf Full one-loop corrections to the relic density in the
MSSM: A few examples}}

\vspace{8mm}

{\large N. Baro${}^{1)}$, F. Boudjema${}^{1)}$ and A.~Semenov${}^{1,2)}$ }\\

\vspace{4mm}

{\it 1) LAPTH$^\dagger$, B.P.110, Annecy-le-Vieux F-74941, France}
\\ {\it
2) Joint Institute of Nuclear Research, JINR, 141980 Dubna, Russia }\\

\vspace{10mm}

\end{center}

\centerline{ {\bf Abstract} } \baselineskip=14pt \noindent

{\small We show the impact of the electroweak, and in one instance
the QCD, one-loop corrections on the relic density of dark matter
in the MSSM which is provided by the lightest neutralino. We cover
here some of the most important scenarii: annihilation into
fermions for a bino-like neutralino, annihilation involving gauge
bosons in the case of a mixed neutralino, the neutralino-stau
co-annihilation region and annihilation into a bottom quark pair.
The corrections can be large and should be taken into account in
view of the present and forthcoming increasing precision on the
relic density measurements.  Our calculations are made  possible
thanks to a newly developed automatic tool for the calculation at
one-loop of any process in the MSSM. We have implemented a
complete on-shell gauge invariant renormalisation scheme, with the
possibility of switching to other schemes. In particular we will
report on the impact of different renormalisation schemes for
$\tan \beta$.}

\vspace*{\fill}

\vspace*{0.1cm} \rightline{LAPTH-1211/07}

\vspace*{1cm}

$^\dagger${\small UMR 5108 du CNRS, associ\'ee  \`a
l'Universit\'e de Savoie.} \normalsize

\vspace*{2cm}

\end{titlepage}

\renewcommand{\topfraction}{0.85}
\renewcommand{\textfraction}{0.1}
\renewcommand{\floatpagefraction}{0.75}
\newcommand{\drbar}{{\overline{\rm DR}}}

\section{Introduction}
The last few years have witnessed spectacular advances in
cosmology and astrophysics confirming with an unprecedented level
of accuracy that ordinary matter is a minute part of what
constitutes the Universe at large. At the same time as the LHC
will be gathering data, a host of non collider experiments will be
carried out in search of Dark Matter, DM,  (either direct or
indirect) as well as through ever more precise determination of
the cosmological parameters. In this new paradigm, the search for
DM at the LHC is high on the agenda, as is of course the search
for the Higgs. In fact these may be two facets of the New Physics
that provides a resolution to the hierarchy problem posed by the
Higgs in the Standard Model, \smp. The epitome of this New Physics
is supersymmetry which among many advantages furnishes a good DM
candidate through the lightest neutralino, $\neuto$. If future
colliders discover supersymmetric particles and probe their
properties, one could predict the dark matter density of the
Universe and would constrain  cosmology with the help of precision
data\cite{wmaplhclc-requirements,Peskin-DM-requirements} provided
by WMAP\cite{wmap} and PLANCK\cite{planck}. It would be highly
exciting if the {\em precision} reconstruction of the relic
density from observables at the colliders does not match PLANCK's
determination, this would mean that the  post-inflation era is
most probably not entirely radiation
dominated\cite{nonconventional-relic}. Already now the accuracy on
the relic dark matter density is about $10\%$ from WMAP and will
soon be improved to about $2\%$ from PLANCK. Such level of
accuracy must be matched by precise theoretical calculations. From
the particle physics point of view this means precision
calculations of the annihilation and co-annihilation cross
sections at least at one-loop. Quite sophisticated codes now
exist\cite{micromegas,Darksusy} for the calculation of the relic
density, however they are essentially based on tree-level cross
sections with the inclusion of some higher order effects
essentially through some running couplings, masses or some
effective couplings. Some of these corrections\cite{micromegas}
have already been shown to be essential like the corrections to
the Higgs couplings that can completely change the picture in the
so-called Higgs funnel (annihilation mainly through the
pseudo-scalar Higgs, $A$). The use of other approximations needs
to be justified by complete higher order calculations which
contain more than just the effect of effective couplings. In a
word, the level of accuracy that will soon be reached requires
that one is ready to tackle in a general way  a full calculation
at one-loop for any annihilation (or co-annihilation) of the
neutralinos in supersymmetry, just as what one has been doing for
the cross sections at the colliders. \\
\noi The aim of this letter is to report on the progress towards
automatisation of these calculations and to show and discuss some
results on the one-loop corrected annihilation and co-annihilation
cross sections of the LSP neutralino in the MSSM. In particular,
we study here some of the most important scenarii: i) annihilation
in the so-called bulk region into fermions for a bino-like
neutralino, ii) co-annihilation involving the neutralino and the
lightest stau $\stauo$, iii) annihilation into a pair of massive
gauge bosons in the case of a mixed neutralino and iv)
annihilation into $b \bar{b}$ where the pseudo-scalar Higgs pole can play a role.
We concentrate on the electroweak corrections, although iv) is an
excuse to show how we handle some classes of QCD corrections.\\
The couple of very recent calculations of loop corrections to the
relic density tackled either QCD corrections in extreme, though
highly interesting, scenarii such as annihilation into top quarks
at threshold\cite{lsptott-rc} and the nice study of
stop-neutralino
co-annihilation\cite{Freitas-relic-qcd}\footnote{We do not list
here loop induced annihilation processes such $\neuto \neuto \to
gg$\cite{lsptogg,sloopsgg,lsplsp-ffg,Barger-lsp-nlo}. A very
recent paper discusses the QCD correction to annihilation into
$b\bar b$ in the funnel\cite{Klasen-relic-qcd}, however the bulk
of all contributions has been known for sometime and implemented
in {\tt micrOMEGAs} already.}. Some important non-perturbative
electroweak effects of the Coulomb-Sommerfeld type that occur for
TeV winos or higgsinos with a relative mass splitting between the
lightest supersymmetric particle (LSP) and the next-to-lightest
supersymmetric particle (NLSP) of $\ord(10^{-4})$ have been
reported in\cite{Nojiri-gammaray-coulomb,Hisano-relic}. Let us
also note that, though not to be seen as radiative corrections to
the annihilation cross sections, the temperature corrections to
the relic density have been considered and found to be totally
negligible at the level of $10^{-4}$\cite{relic-temp-corr}. A
better simulation of the cosmological equation of state to derive
the effective number of relativistic degrees of freedom has been
done giving corrections ranging from $1.5\%$ to
$3.5\%$\cite{Hindmarsh-gstar} compared to the usual treatment as
done in {\tt DarkSUSY} \cite{Darksusy} or {\tt
micrOMEGAs}\cite{micromegas}.

\section{General set-up and details of the calculation}
\subsection{Set-up of the automatic calculation of the cross sections}
Even in the \smp, one-loop calculations of $2\ra 2$ processes
involve hundreds of diagrams and a hand calculation is practically
impracticable. Efficient automatic codes for any generic $2\ra 2$
process, that have now been exploited for many $2\ra
3$\cite{grace2to3,other2to3} and even some $2\ra
4$\cite{grace2to4,Dennereeto4f} processes, are almost unavoidable
for such calculations. For the electroweak theory these are the
{\tt GRACE-loop}\cite{grace-1loop} code and the bundle of packages
based on  {\tt FeynArts}\cite{FeynArts}, {\tt
FormCalc}\cite{FormCalc}
and {\tt LoopTools}\cite{looptools}, that we will refer to as {\tt FFL} for short. \\
\noindent With its much larger particle content, far greater
number of parameters and more complex structure, the need for an
automatic code at one-loop for the minimal supersymmetric standard
model is even more of a must. A few parts that are needed for such
a code  have been developed based on an extension of
\cite{FeynArtsusy} but, as far as we know, no complete code exists
or is, at least publicly, available. {\tt
Grace-susy}\cite{Grace-susy} is now also being developed at
one-loop and many results exist\cite{Grace-susy1loop}. One of the
main difficulties that has to be tackled is the implementation of
the model file, since this requires that one enters the thousands
of vertices that define the Feynman rules. On the theory side a
proper renormalisation scheme needs to be set up, which then means
extending many of these rules to include counterterms. When this
is done one can just use, or hope to use, the machinery developed
for the \smp, in particular the symbolic manipulation part and
most importantly the loop integral routines including tensor
reduction algorithms or any other efficient set of basis integrals.\\
\noindent The results we will report are based on the development of a new automatic
tool that uses and adapts modules, many of which, but not all, are
part of other codes like {\tt FFL}.  This is the package {\tt
SloopS} whose main components and architecture we briefly sketch.

In this application we combine {\tt LANHEP}\cite{lanhep}
(originally part of the package {\tt COMPHEP}\cite{comphep}) with
the {\tt FFL} bundle but with an extended and adapted {\tt
LoopTools}\cite{sloopsgg}. {\tt LANHEP} is a very powerful routine
that {\em automatically} generates all the sets of  Feynman rules
of a given model, the latter being defined in a simple and compact
format very similar to the canonical coordinate representation.
Use of multiplets and the superpotential is built-in to minimize
human error. The ghost Lagrangian is derived directly from the
BRST transformations. The {\tt LANHEP} module also allows to shift
fields and parameters and thus generates counterterms most
efficiently. Understandably the {\tt LANHEP} output file must be
in the format of the model file of the code it is interfaced with.
In the case of {\tt FeynArts} both the {\it generic} (Lorentz
structure) and {\it classes} (particle content) files had to be
given. Moreover because  we use a non-linear gauge fixing
condition\cite{grace-1loop}, see below, the {\tt FeynArts} default
{\it generic} file had to be extended.

\subsection{Renormalisation and renormalisation schemes}
\label{subsec:renorm} In the last half decade there has been an
upsurge and flurry of activity constraining models of
supersymmetry and  other New Physics with the limit on the relic
density delimiting most of the parameter space of these models.
All these investigations are based on tree-level, sometimes with
improved effective couplings, estimates of the relic density. Only
in the last few months have some
investigations\cite{Leszek-deltaom}, within mSUGRA, added a
theoretical error estimate of $\ord(10\%$), {\it i.e}, of the same
order as the current experimental error. In these analyses based
on renormalisation group running, a substantial uncertainty is due
to the impact of the running
itself\cite{wmaplhclc-requirements,gene-relic-rge}. Even if the
weak scale spectrum is known, loop corrections to the cross
sections are needed. In fact the precision one-loop calculations
we are carrying will be most useful when confronting a measurement
of the relic density once the microscopic properties of dark
matter would have been pinned down at the collider and in
direct/indirect detection. Henceforth we   rely on the {\em
physical masses} of the SUSY particles with the addition of some
{\em physical observables} to fully reconstruct the model. We
therefore work, as far as possible, within an on-shell scheme
generalising what is done for the electroweak standard model.\\
\noi i) The Standard Model parameters: the fermion masses as well
as the mass of the $W$ and the $Z$ are taken as input physical
masses. The electric charge is defined in the Thomson limit, see
for example\cite{grace-1loop}. Because we are calculating
corrections to processes at a scale $2\mneuto \sim 2 M_Z$, the
effect from the running electromagnetic coupling due to the light
fermion masses will, alone, rescale the tree-level cross section
leading to a correction of about $15\%$ to the cross sections. The
light quark (effective) masses, are chosen such as to reproduce
the \sm value of $\alpha^{-1}(M_Z^2)=127.77$ including  the light
fermions contribution, which give the $\sim 15\%$ corrections
compared to the use of $\alpha(0)$. For the \sm input masses see
the last papers of Ref.~\cite{grace2to3} with the exception of
$m_{\rm top}=170.9\;$GeV. We will keep this rescaling in mind.
This effect can be reabsorbed by using a scheme where the
effective $\alpha(M_Z^2)_{{\rm eff.}}$ is used as
input. \\
\noi ii) The Higgs sector: The Higgs sector is conceptually the
trickiest. First we take $M_A$ the pseudoscalar Higgs mass as an
input parameter and require vanishing tadpoles. The extraction and
definition of the ubiquitous $\tgb$, which at tree-level is
identified as the ratio of  the $vev$ of the two Higgs doublet is
the tricky part. Most schemes define the $\tgb$ counterterm at
one-loop from a non physical quantity, such as the $A^0Z$
transition two-point function at $q^{2}=M_A^{2}$.  It has become
customary to take a $\overline{DR}$ definition, by only taking
into account the ``universal" ultraviolet part from such
quantities,  leaving out all finite parts. These prescriptions are
however not gauge invariant, see for
example\cite{Freitas-Stockinger-tb}. Moreover the ``universal"
part is only universal in the usual linear gauge. With the
non-linear gauge fixing we implement, see
section~\ref{subsec:nlg}, our results would not be gauge invariant
and one has to be very careful with the Ward identities. We leave
this important issue to a forthcoming
paper\cite{Sloops-higgspaper}. Nonetheless to conform with this
widespread general usage, we also implement a $\overline{DR}$
scheme defined from a physical quantity, to be discussed shortly,
which  reproduces the usual counterterms defined from other
quantities in the linear gauge. As known, the others Higgs masses
$m_{h^{0}}$ (for the lightest CP-even), $m_{H^{0}}$ (for the
heaviest CP-even) and $m_{H^{\pm}}$ (for the charged) receive
corrections that can be very important. To be able to stick with
an on-shell definition and in order to weigh the effect of the
$\tgb$ scheme dependence, we also define two other schemes. One is
based on the use of $A^{0} \ra \tau^+ \tau^-$ partial width to
which the QED corrections have been extracted, we will refer to
this scheme as the $A_{\tau \tau}$ scheme. For the third one, we
take $m_{H^{0}}$ as an input parameter and trade it for ``$\tgb$"
hence loosing one prediction, we will call this scheme $MH$. This
scheme is also used in\cite{Grace-susy1loop}. With
\tgbt fixed, we can turn to the other sectors. \\
\noi iii) Neutralino and charginos: For the neutralino and
chargino sector, we implement an on-shell scheme taking as input
parameters the masses of the two charginos (this defines the
counterterms to the SU(2) gaugino, wino $\tilde w$, mass $M_2$ and
to the higgsino, $\tilde h$, parameter $\mu$) and the mass of the
LSP $\mneuto$ (which completes the extraction of the  U(1)
gaugino, bino $\tilde b$, mass $M_1$). The other neutralino masses
$m_{\chi_{2}^{0}}$, $m_{\chi_{3}^{0}}$ and $m_{\chi_{4}^{0}}$
receive corrections to their tree-level value. Obtaining finite
corrections for the masses and decays is a not trivial test of the
procedure. Here our implementation is quite similar to the one
adopted in\cite{Hollik-chargino-rc} when one
sticks to the $\overline{DR}$ \tgbt. \\
\noi iv) Sfermions: For the slepton sector we use as input parameters the physical
masses of the two charged sleptons which in the case of no-mixing
define the $R$-slepton soft breaking mass, $\tilde{M}_{\tilde{l}_
R}$ and the $SU(2)$ mass, $\tilde{M}_{\tilde{l}_L}$, giving a
correction to the sneutrino mass at one-loop. In the case of mixing one needs
to fix the counterterm to the tri-linear coupling. The best option
would have been to define this from a decay such as $\staut \ra
\stauo Z$. In the present letter we take a much simpler
prescription, we solve the system by taking as input all three
slepton masses.
For the squark sector, for each generation three  physical masses
are needed as input to constrain  the breaking parameters
$\tilde{M}_{\tilde{Q}_{L}}$ for the $SU(2)$ part,
$\tilde{M}_{\tilde{u}_{R}}$, $\tilde{M}_{\tilde{d}_{R}}$ for the
$R$-part. In case of mixing, the simplest prescription for the
counterterms to the tri-linear couplings $A_{u}$, $A_{d}$ derives
from two conditions on the renormalised mixed two-point functions
as is done in\cite{Hollik-sfermion-rc}. Our plan is to replace
these conditions by an on-shell input such as a decay of the heavy
squark to the lighter one and a $Z$, to conform with a fully
on-shell scheme and study further the scheme dependence. \\
\noi Wave function renormalisation is introduced so that the residue at
the pole of all physical particles is $1$ and no-mixing is left
between the different particles when on shell. This applies for
all sectors. Dimensional reduction is used as implemented in the
{\tt FFL} bundle at one-loop through the equivalent constrained
dimensional renormalisation\cite{CDR}. Renormalisation of the
strong coupling constant and the gluino is not an issue for the
examples we study here.

We have verified our codes and schemes with different calculations
on the market for a variety of correction to masses and other
observables\cite{Grace-susy,Hollik-sfermion-rc,Hollik-chargino-rc,
Freitas-Stockinger-tb}. The code has also been used for
corrections to \sm processes and also to one-loop induced
processes $\neuto \neuto \ra \gamma \gamma, Z\gamma,
gg$\cite{sloopsgg} relevant for indirect detection.

\subsection{Non-Linear gauge-fixing}
\label{subsec:nlg} We use a generalised non-linear
gauge\cite{nlg-generalised,sloopsgg} adapted to the minimal
supersymmetric model. The gauge fixing writes
\begin{eqnarray}
\label{gaugefix} {\mathcal L}_{GF}& = &-
\frac{1}{\xi_W}|(\partial_\mu - ie
  \tilde \alpha \gamma_\mu - igc_W \tilde \beta
  Z_\mu) W^{\mu \, +} + \xi_W \frac{g}{2}(v + \tilde \delta h + \tilde
  \omega H + i \tilde \kappa G_3 + i \tilde \rho A) G^+|^2 \nonumber \\
&-& \frac{1}{2\xi_Z} (\partial . Z + \xi_Z \frac{g}{2c_W}(v +
  \tilde \epsilon h + \tilde \gamma H) G_3)^2  -
  \frac{1}{2\xi_\gamma} (\partial . \gamma)^2.
\end{eqnarray}
Unlike the other parts of the model, ${\mathcal L}_{GF}$ is
written in terms of {\em renormalised} fields and parameters.
$G_3,G^\pm$ are the Goldstone fields. We always work with
$\xi_{\gamma,Z,W}=1$ so as to deal with the minimal set of loop
tensor integrals. More details will be given
elsewhere\cite{Sloops-higgspaper}.

\subsection{The different parts of the cross section}
The one-loop amplitudes consist of the virtual corrections
$\mathcal{A}_{1loop}^{EW+QCD}$ and the counterterm contributions
$\mathcal{A}_{CT}$. $\mathcal{A}_{1loop}^{EW+QCD}+
\mathcal{A}_{CT}$ must be ultraviolet finite. On the other hand
$\mathcal{A}_{1loop}^{EW+QCD}$ can contain infrared divergences
due to photon and gluon virtual exchange. These are regulated by a
small photon or gluon mass. For  the QCD corrections we study
here, this implementation does not pose a problem. The photon and
gluon mass regulator contribution contained in the virtual
correction should cancel exactly against the one present in the
photon and gluon final state radiation. The photonic (gluonic)
contribution is in fact split into a soft part, where the photon
(gluon) energy is less than some small cut-off $k_c$,
$\mathcal{A}_{\gamma,g}^{soft}(E_{\gamma,g}<k_c)$ and a hard part
with $\mathcal{A}_{\gamma,g}^{hard}(E_{\gamma,g}>k_c)$. The former
requires a photon/gluon mass regulator. We use the usual universal
factorised form with a simple rescaling for the case of the gluon
correction. We take $\alpha_s=\alpha_s(M_Z^2)=0.118$.

\subsection{Checks on the calculation}

\noi {\it {\bf i)}} For each process and for each set of parameters,
we first check the ultraviolet finiteness of the results. This
test applies to the whole set  of the virtual one-loop diagrams.
The ultraviolet finiteness test is performed by varying the
ultraviolet parameter $C_{UV}=1/\varepsilon$. We vary
$C_{UV}$ by seven orders of magnitude with no change in the
result. We content ourselves with double precision.

\noi {\it {\bf ii)}} The test on the infrared finiteness is
performed by including both the loop and the soft bremsstrahlung
contributions and checking that there is no dependence on the
fictitious photon mass $\lambda_{\gamma}$ or gluon mass
$\lambda_g$.

\noi {\it {\bf iii)}} Gauge parameter
independence of the results is essential. It is performed through
the set of the {\em seven} gauge fixing parameters defined in
Eq.~(\ref{gaugefix}). The use of the seven parameters is not
redundant as often these parameters check complementary sets of
diagrams. It is important  to note that in order to successfully
achieve this test one should not include any width in the
propagators. In fact our tree-level results do not include any
width. Because of the parameters and the energies we consider, no
width is required to regularise the cross sections.

\noi {\it {\bf iv)}} For the bremsstrahlung part  we use VEGAS
adaptive Monte Carlo integration package provided in the {\tt FFL}
bundle and verify the result of the cross section against {\tt
CompHep} \cite{comphep}. We choose $k_c$ small enough and  check
the stability and independence of the result with respect to
$k_c$.

\subsection{Boltzmann equation, the small $v$ expansion}
Having the collection of cross sections and the masses of the
annihilating (and co-annihilating) DM particles  we could have
passed them to {\tt micrOMEGAs} for a very precise determination
of the relic density based on a careful treatment of the Boltzmann
equation. However, to weigh the impact of the corrections on the
relic density it is worth to gain insight through an approximation
in going from the cross sections to the relic density, especially
that we have found these approximations to be, after all, rather
excellent for the cases we study, including co-annihilations.
Moreover corrections to the cross sections could be incorporated
in the case of non-thermal production. All cross sections
$\sigma_{ij}$ where $i,j$ label the annihilating and
co-annihilating DM particles $i,j$, are expanded in terms of the
relative velocity $v_{ij}$, which for neutralino annihilation is
$v=2\beta=2\sqrt{1-4\mneuto^2/s}$. Away from poles and thresholds,
it is a very good approximation to write
$\sigma_{ij}v_{ij}=a_{ij}+ b_{ij} v^2$, keeping only the $s$-wave,
$a_{ij}$, and $p$-wave, $b_{ij}$ coefficients. With $T$ being the
temperature, $x=\mneuto/T$, the thermal average gives
\beqn
\label{thermalav}
 \langle \sigma_{ij} \; v_{ij} \rangle=a_{ij}+6
(b_{ij}-a_{ij}/4)/x,
\eeqn
with $g_1=2$ the neutralino spin degree of freedom (sdof), the
co-annihilating particle of sdof $g_i$ and mass $m_i$ contributes
an effective relative weight of
\beqn
\tilde{g}_{i,eff}=\frac{g_i}{g_1} (1+\delta_i)^{3/2} \exp (-x
\delta_i), \quad \delta_i=(m_i-\mneuto)/\mneuto.
\eeqn
\noi The total number of sdof is $\tilde{g}_{eff}=\sum_i
\tilde{g}_{i,eff}$. A good approximation for the relic density is
obtained by carrying a simple one dimensional integration
\beqn
\label{relic-app-coann} \Omega h^{2} &=&
\left(\frac{10}{\sqrt{g_*(x_F)}} \; \frac{x_F}{24} \;  \right)
\frac{0.237 \times 10^{-26} {\rm cm}^{3}.{\rm s}^{-1}}{x_F \; J},
\quad
J= \int_{x_F}^{\infty } \langle \sigma v \rangle_{eff} dx/x^2 \nonumber \\
\langle \sigma v \rangle_{eff}&=&\sum_{ij} \frac{\tilde{g}_{i,eff}
\tilde{g}_{j,eff}}{\tilde{g}_{eff}^2}\; \langle \sigma_{ij} \;
v_{ij} \rangle.
\eeqn
$a_{ij}, b_{ij}$ that are needed to compute $\sigma_{ij}$ in
Eq.~(\ref{relic-app-coann}) are given in ${\rm cm}^{3}.{\rm
s}^{-1}$. $x_F$ represents the freeze-out temperature. $g_*(x_F)$
is the effective degrees of freedom at freeze-out. $g_*$ is
tabulated in {\tt micrOMEGAs}  and we read it from there. For the
examples we will study $\sqrt{g_*(x_F)}\sim 9.29$. In the
freeze-out approximation, $x_F$ can be solved iteratively from
\begin{eqnarray}
\label{xf-solve} x_{F}=21.2181 + \ln \left( \frac{(\tilde{g}_{eff}
\langle \sigma v \rangle_{eff})|_{x_F}}{10^{-26}}
\frac{\mneuto}{100} \sqrt{{\frac{2400}{g_*(x_F)  x_F}}} \; c(c+2)
\right),
\end{eqnarray}
where the neutralino mass is expressed in GeV. The numerical
solutions of the density equation and hence the freeze-out
suggest\cite{Gelmini-Gondolo,micromegas} $c=1.5$ is a very good
choice in most, but not all, cases. Though we have verified that
Eq.~(\ref{xf-solve}) converges quickly and agrees well with the
result of {\tt micrOMEGAs}, the results we give use $x_F$ as
extracted from {\tt micrOMEGAs}. The loop corrected cross sections
should also impact on the value of $x_F$ which is  not  exactly
the same as the value extracted from the tree-level cross
sections. However, the shift is marginal, though ultimately in a
full computation this should be taken into account. Our results
therefore use the same value of $x_F$ at both tree and one-loop
level. On the other hand to derive the relic density we rely, in
this letter, on Eq.~(\ref{relic-app-coann}). For The case of $\neuto
\neuto$ annihilation, the latter simplifies to
\beqn
\label{relic-app-ann} \Omega h^{2} &=&
\left(\frac{10}{\sqrt{g_*(x_F)}} \; \frac{x_F}{24} \;  \right)
\frac{0.237 \times 10^{-26} {\rm cm}^{3}.{\rm
s}^{-1}}{a+3(b-a/4)/x_F}.
\eeqn
\noi The weight of a channel (see the percentages we will refer to
later) corresponds to its relative contribution to $J$.
\subsection{Choosing points in the MSSM parameter space}
Current limits on the relic density, from WMAP and SDSS\cite{wmap}
give the $2\sigma$ range
\beqn
0.092 < \Omega h^2 < 0.124.
\eeqn
In this first exploratory study we thought it is best  to consider
different scenarii without worrying too much about the absolute
value of the derived relic density in order to grasp the origin of
the large corrections, if any. Our choice of scenarii was
motivated by the physics issues, although our choice is biased
towards the popular scenarii that emerge in mSUGRA within thermal
production. Nonetheless our choice covers annihilation into
fermions, gauge bosons and co-annihilation. This said, apart from
the annihilation into gauge bosons, the derived relic density is
either within this range or not overly outside. For the gauge
bosons the motivation was to take a model that singles out the
$WW$ and $ZZ$ final states channels. Moreover since the impact of
the radiative corrections can be large there is not much sense in
picking up a model based on its agreement with the current value
of the relic density on the basis of a tree-level calculation.
This said we have used {\tt micrOMEGAs} as a guide, being careful
about translations of effective couplings and input parameters.
{\tt micrOMEGAs} was also quite useful in justifying the
approximations we use for deriving the relic density from the
cross sections. We should also add that in this letter we do not
apply the radiative corrections to all the subprocesses that can
contribute to the relic density but only to those channels and
subprocesses that contribute more than $5\%$ to the relic density.
When calculating the correction to the relic density we include
these channels at tree-level.


\section{Annihilation of a bino LSP into charged
leptons} \label{sec-bulk}

 The first example we take corresponds to
the so-called bulk region, with a neutralino LSP which is mostly
bino. The latter will couple with the particles of largest
hypercharge, the R-sleptons. Therefore annihilation is into
charged leptons. Because of the Majorana nature of the LSP, there
is no $s$-wave contribution in the case of massless fermions. In
our case the contribution to the $s$-wave (at tree-level) is from
the $\tau$'s. In the radiation dominated standard scenario, to be
consistent with the present value of the relic density , we take
right sleptons as light as possible (but within the LEP limits)
while all other particles (squarks, charginos, other neutralinos)
heavy. The example we take has $\tgb=5$. The relevant physical
parameters are $\mneuto=90.72\;$GeV$, \mstauo=115.15\;$GeV and
$\mser=\msmur=117.5\;$GeV. The masses of the charginos are
$\mchargo=200.64,\mchargt=610.47\;$GeV. To give an idea this
reconstructs the {\it tree-level} values for the gaugino and
higgsino parameters as $M_1=90\;$GeV while $M_2=200,\mu=-600\;$GeV
leading to a neutralino which is almost $100\%$ bino: $\neuto=
0.998\tilde{b} +0.012 \tilde{w} -0.068\tilde{h}_1
+0.003\tilde{h}_2$.

The other physical input masses reconstruct the tree-level
parameters $m_{\tilde e_R}=\;$110GeV, $\tilde M_{\tilde
E_L}=250\;$GeV, $M_3=\tilde M_{\tilde Q_L}=m_{{\tilde u,\tilde
d}_R}=800\;$GeV, $M_A=500\;$GeV and the tri-linear coupling
$A_f=0$. The contribution to the relic density is, then as
expected, into leptons $(98\%$) with the proportions as shown in
Table~\ref{tab-bino}. The $e^{+}e^{-}$, $\mu^{+}\mu^{-}$ channels
contribute for $31\%$ each. The difference between the three
channels is accounted for by the contribution of the $s$-wave of
the $\tau$ final states and very little from the fact that
$\tilde{\tau}_1$ is slightly lighter than the lightest smuon and
selectron.
\begin{table}
\begin{center}
\begin{tabular}{|c|c|c|c|c|c|}
\hline $\neuto\neuto\rightarrow \tau^+ \tau^-$ ($36\%$)&  Tree & & $A_{\tau \tau}$ & $\drbar$ & $MH$ \\
\hline a & 0.081 & & +38\% & +35\% & +15\% \\
\hline b & 3.858 & & +18\% & +18\% & +18\% \\
\hline
\hline $\Omega h^{2}$ & 0.166 & & 0.138 & 0.138 & 0.141 \\
\hline $\frac{\delta \Omega h^{2}}{\Omega h^{2}}$ &  & & -17$\%$ & -17$\%$ & -15$\%$\\
\hline
\end{tabular}
\caption{{\em Bino case: Tree-level values of the $s$-wave ($a$)
and $p$-wave ($b$) coefficients in units $ 10^{-26} {\rm cm}^{3}
{\rm s}^{-1}$, as well as the relative one-loop corrections in the
$\drbar$, $A_{\tau \tau}$ and $MH$ scheme. The derived relic
density and its percentage change are also given. ($36\%$) to the
annihilation process refers to the percentage weight, at
tree-level, of the channel}. \label{tab-bino} }
\end{center}
\end{table}

Let us first comment on the $p$-wave contribution which gives the
bulk of the contribution to the relic density. The total
correction is about $18\%$ in this case. It is tempting to
parameterize the corrections. In fact, had we used the value of
the gauge coupling not at low energy but at the scale of the
$\neuto$ mass of order $M_Z$ the bulk of the correction would be
absorbed. Indeed, $(137.04/127.77)^2-1 \sim 15\%$. In the few
other examples we have looked at concerning the annihilation into
leptons, we arrive at the same order of correction, see for
example the corrections to the $p$-wave in the case of
co-annihilation and even where there is some higgsino component as
in the case of annihilation into $b \bar{b}$. The other common
trend is that the correction does not show any dependence on the
\tgbt scheme we choose when there is a large bino component. This
is not the case for the $s$-wave. In particular the $MH$ scheme
differs from the $\drbar$ and $A_{\tau \tau}$. The \tgbt
dependence comes essentially from the Yukawa contribution, see
Eq.~(4.8) of ref\cite{wmaplhclc-requirements}. The latter is also
sensitive to the higgsino component of the neutralino that is also
affected by the \tgbt change. The effect is more obvious in the
case of scenario iii), see Table~\ref{tab-mix2}. Note however that
even in the case of massless fermions, there is (though small) a
contribution to the $s$-wave due to hard photon radiation.   Hard
photon radiation in association with a light charged fermion pair
is not subject, for the $s$-wave amplitude, to  the known helicity
suppression when no photon is emitted\cite{lsplsp-ffg}. Taking
both the $s$ and $p$-wave contribution leads to a correction on
the relic density of about $17\%$. As discussed a few lines above,
using $\alpha_{eff}(M_Z)$ reduces the correction to the level of a
few percent.

\section{Neutralino $\tilde{\tau}$ co-annihilation}
\label{sec:coan}

In this scenario the LSP is still the lightest neutralino and we
take it to be essentially bino though with a small higgsino
component with a composition $\neuto= 0.986\tilde{b} -0.049
\tilde{w} +0.144\tilde{h}_1 -0.070\tilde{h}_2$ and mass
$\mneuto=162.34\;$GeV. We consider a scenario where the NLSP is
the lightest $\tilde{\tau}_1$ with mass
$m_{\tilde{\tau}_1}=168.42\;$GeV coming mainly from its
$\tilde{\tau}_R$ component. The lightest smuon and selectron are
given masses so that they are thermodynamically irrelevant, they
have a mass of about $195\;$GeV. Apart from $M_{A}=1$TeV, for the
squark and Higgs sector, as well as \tgbt, the parameters are the
same as in the example in the bulk region. We want therefore to
concentrate on co-annihilation involving only $\tilde{\tau}_1$.
With a mass difference between the LSP and NLSP of $6.08\;$GeV
which is about only $4\%$ of the mass of the LSP,
$\neuto\tilde{\tau}_1$ annihilation is quite efficient with the
two channels $\neuto\tilde{\tau}_1^\pm \ra \tau^\pm
\gamma,\tau^\pm Z $ accounting for as much as half of the total
contribution, see Table~\ref{tab-coann}.
$\tilde{\tau}^\pm_1\tilde{\tau}^\pm_1$ takes up a quarter of the
total. Neutralino annihilation makes up for about $15\%$. The rest
is made up by $\tilde{\tau}^-_1\tilde{\tau}^+_1 \ra  \gamma
\gamma, \gamma Z$. It is interesting to note that
$\tilde{\tau}^\pm_1\tilde{\tau}^\pm_1 \gg
\tilde{\tau}^-\tilde{\tau}^+$. Our approximation based on
Eq.~(\ref{relic-app-coann}) gives $\Omega h^{2}=0.128$, with
$x_F=26.5$.

\begin{table}[h]
\begin{center}
\begin{tabular}{|c|c|c|c|c|c|}
\hline $\neuto\neuto\rightarrow \tau^+ \tau^-$ ($6\%$) &  Tree & & $A_{\tau \tau}$ & $\drbar$ & $MH$ \\
\hline a & 0.002 & & +200\% & +200\% & +200\%  \\
\hline b & 1.717 & & +18\% & +19\% & +19\% \\
\hline
\hline $\neuto\tilde{\tau}^\pm_1 \rightarrow \tau^\pm \gamma$ ($37\%$) &  Tree & & $A_{\tau \tau}$ & $\drbar$ & $MH$ \\
\hline a & 4.342 & & +9\% & +9\% & +9\% \\
\hline b & -1.116 & & +9\% & +8\% & +9\% \\
\hline
\hline $\neuto\tilde{\tau}^\pm_1 \rightarrow \tau^\pm Z$ ($10\%$) &  Tree & & $A_{\tau \tau}$ & $\drbar$ & $MH$ \\
\hline a & 1.093 & & +21\% & +21\% & +21\% \\
\hline b & -0.214 & & +19\% & +19\% & +19\% \\
\hline
\hline $\tilde{\tau}^\pm_1\tilde{\tau}^\pm_1\rightarrow \tau^\pm \tau^\pm$ ($23\%$) &  Tree & & $A_{\tau \tau}$ & $\drbar$ & $MH$ \\
\hline a & 43.345 & & +17\% & +17\% & +17\% \\
\hline b & -14.445 & & +13\% & +13\% & +14\% \\
\hline c & 0      & & -0.994 & -0.994 & -0.994\\
\hline
\hline $\Omega h^{2}$ & 0.128 & & 0.117 & 0.117 & 0.117\\
\hline $\frac{\delta \Omega h^{2}}{\Omega h^{2}}$ &  & & -9$\%$ & -9$\%$ & -9$\%$\\
\hline
\end{tabular}
\caption{{\em Neutralino $\stauo$ co-annihilation: meaning of the
different cells is as in Table~\ref{tab-bino}. Percentages in the
first column represent the weight of the corresponding channels.}
\label{tab-coann} }
\end{center}
\end{table}
$\bullet \; \neuto \neuto \ra f \bar f$\\
\noi  Compared to the bino-case in the bulk region where this
channel accounted for the totality of the relic density, here it
only makes up for $6\%$. Nonetheless the effect of radiative
corrections on this channel are very similar to what we found in
the scenario of section~\ref{sec-bulk}. One may be misled by
interpreting the $200\%$ relative correction as a large correction
to the relic density. This is a relatively large correction to the
$s$-wave contribution, but in absolute terms this correction is
totally negligible compared to the correction brought about by the
$p$-wave contribution at the cross section level as well as after
taking the thermal average, notwithstanding that the whole channel
in the co-annihilation region has little weight which further
dilutes the effect of such large relative correction. As pointed
out in the previous section this $200\%$ relative correction is
due to the smallness of the $s$-wave $\neuto \neuto \ra f \bar f$
which is offset  by the hard photon emission that now allows for
an $s$-wave contribution\cite{lsplsp-ffg}. As discussed in the
first reference in\cite{lsplsp-ffg}, the relative importance of
hard radiation reduces fast as the mass of the intermediate
slepton increases. This explains why the relative effect is more
prominent in the co-annihilation region.

Before getting into the details about each of the main
contributing channels that  involve $\tilde{\tau}^\pm_1$
co-annihilation, let us point at some of their common features.
The bulk of the contribution comes now from the $s$-wave,
especially after taking into account thermal averaging. Another
common feature is that the \tgbt scheme dependence is hardly
noticeable. Moreover, the corrections to the $s$- and $p$-wave
are, within a margin of $4\%$, the same.

\noi $\bullet \; \neuto\tilde{\tau}^\pm_1 \rightarrow \tau^\pm \gamma,\tau^\pm Z$\\
For $\neuto\tilde{\tau}^\pm_1 \rightarrow \tau^\pm \gamma$ the
correction to the $p$-wave is within only a $1\%$ from the
correction to the $s$-wave. Here the electroweak correction
amounts to about $9\%$. This is about half the correction we find
in all other channels. The reason is the following, the effective
coupling for the emitted photon should still be taken at $q^2=0$,
and therefore effectively since there is only one neutralino
coupling, this should be taken at the scale of $\mneuto$. Proper
use of the effective couplings here absorbs about $7\%$ of the
correction, leaving therefore a $2\%$ correction. For the $\tau Z$
final state the use of the effective coupling would leave out a
$6\%$ correction to the $s$-wave after absorbing the correction due to
the effective coupling. \\

\begin{figure*}[htbp]
\begin{center}
\includegraphics[width=0.85\textwidth,height=0.6\textwidth]{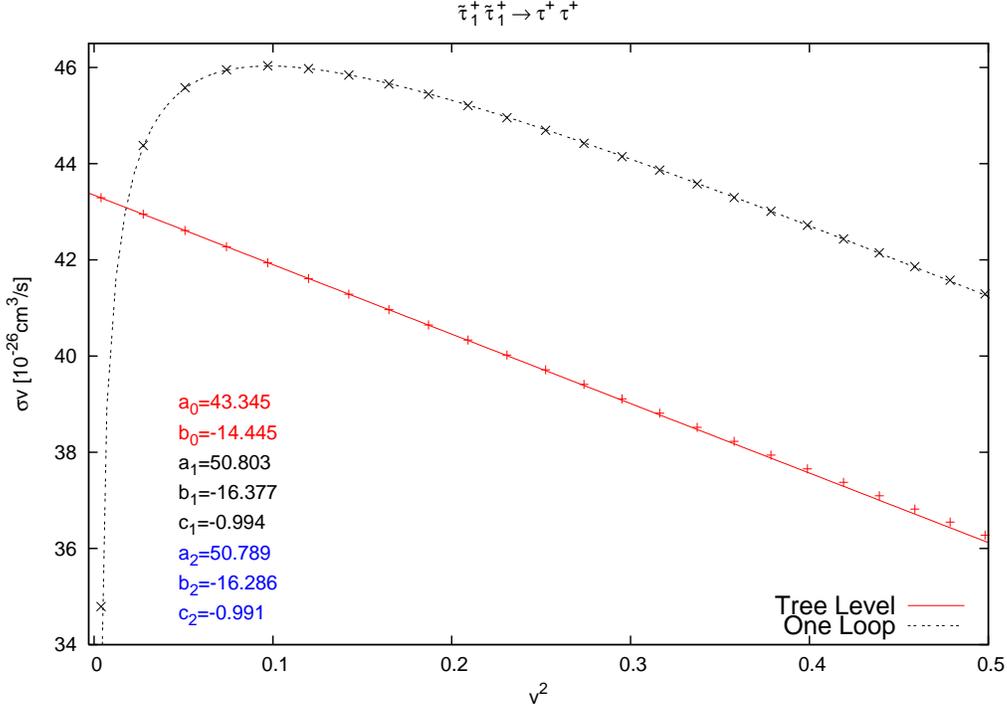}
\caption{\label{fig.coulomb}{\em  $\stauo^\pm \stauo^\pm \ra
\tau^\pm \tau^\pm$ as a function of the square of the relative
velocity both for tree-level (solid line) and at one-loop (dotted
line) in the $A_{\tau \tau}$ scheme. Fits to the $s$-wave ($a$),
$p$-wave ($b$) and the Coulomb factor are given. $a_0,b_0$ for
tree-level, $a_1,b_1$ for the loop with $c_1$ given by
Eq.~(\ref{eq:coulomb}), $a_1,b_1,c_2$ when all three parameters are
extracted from the fit. The two fits at one-loop are not
distinguishable in the figure.}}
\end{center}
\end{figure*}

\noi $\bullet \;  \stauo^\pm \stauo^\pm \ra \tau^\pm \tau^\pm$ \\
\noi The radiative correction to $\stauo^\pm \stauo^\pm \ra
\tau^\pm \tau^\pm$ reveals a quite interesting feature as can be
seen from Fig.~\ref{fig.coulomb} which shows the dependence of the
cross section as a function of the relative velocity (or rather
the square of it). It is from this velocity dependence that we
usually extract the values of the coefficients of the $s$- and
$p$-wave contribution, at both the tree-level and one-loop, that
we need to calculate the relic density.  The figure extends to
$v^2$ values as large as $0.5$ while we  could have contented
ourselves with a maximal value of $v^2=0.3$, considering the
typical value that one obtains for the freeze-out temperature
$\langle v^2 \rangle \sim 6/x_F\sim 0.2$. Even so, the figure
shows that in this case a fit to the tree-level cross section in
the form $a+b v^2$ works quite well. For the one-loop correction a
polynomial fit does not do for low enough velocities. There is a
large negative correction for $v \ra 0$. This correction is in
fact very easy to understand. It is the perturbative one-loop
manifestation of the non relativistic Coulomb-Sommerfeld
effect\cite{Sommerfeld}.
With the tree-level cross section denoted as $\sigma_0$ and
$\sigma_0 v=a_0+b_0 v^2$, the one-loop perturbative cross section
for the same-sign stau annihilation
$\tilde{\tau}^\pm_1\tilde{\tau}^\pm_1$, $\sigma_{{\rm
Coul.}}^{{\rm 1-loop}}$, is such that
\beqn
\label{eq:coulomb0} \frac{\sigma_{{\rm Coul.}}^{{\rm
1-loop}}}{\sigma_0}=-\frac{\pi \alpha}{v}.
\eeqn
We thus expect for the one-loop cross section, $\sigma_1$
\beqn
\label{eq:coulomb} \sigma_1 v=a_1 + b_1 v^2 + c_1/v \quad {\rm
with}\;\; c_1=-\pi \alpha a_0.
\eeqn
Fig.~\ref{fig.coulomb} reflects this repulsive behaviour
perfectly. In fact we made a fit to the one-loop result with the
function $\sigma_1 v=a_1 + b_1 v^2 + c_1/v$, first with  $c_1$ as
given in Eq.~(\ref{eq:coulomb}) and then with $c_1$  not
constrained. The two fits are practically indistinguishable in
Fig.~\ref{fig.coulomb}. Our automatic calculation code captures
this effect perfectly.

One may ask about how to deal with the $1/v$ singularity. In fact
when calculating the relic density, the one-loop $1/v$ singularity
at the level of the cross section  is tamed after thermal average,
$\propto \int_0^{\infty} ({\rm d}v \; v^2 \; e^{-v^2/4 x} ) \;
(\sigma v)$, see also Eq.~\ref{thermalav}. At the end its
contribution to the relic density compared to $a_1$ is
approximately $x_F^{-1}(a_1-2 a_0 \alpha \sqrt{\pi x_F}$).  In
words, non zero temperature of the problem provides a cut-off. One
can also ask whether the one-loop result from the
Coulomb-Sommerfeld effect is sufficient. As seen, the QED
correction is of ${\cal O}(\pi \alpha/v)\sim 0.17$ with $v$
typical of the freeze-out temperature. Therefore in our case a
one-loop treatment seems to be sufficient especially that the
$\stauo^\pm \stauo^\pm \ra \tau^\pm \tau^\pm$ is not the most
dominant channel. This said, these non-relativistic QED $1/v$
threshold corrections can be resummed to all orders. This
resummation as originally performed by Sommerfeld\cite{Sommerfeld}
has been known for quite a long time in quantum mechanics, see
\cite{textbook-Sommerfeld} for a textbook treatment, and amounts
to solving the Schr\"odinger equation in the Coulomb potential.
With $X_{nr}= -2 \pi \alpha/v$ for the same-sign stau
annihilation, the $s$-wave factor resums to
\beqn
\label{resum-Sommerfeld}
S_{nr}=\frac{X_{nr}}{1-e^{-X_{nr}}}=1-\frac{\pi \alpha}{v}+
\frac{1}{3} \left(\frac{\pi \alpha}{v} \right)^2 + \cdots
\eeqn
One might question the validity of Eq.~\ref{resum-Sommerfeld} in
our case where $\stau_1$ is not stable. Finite decay width can of
course act as a cut-off for the $1/v$ corrections, see the case
 of $W$ pair production\cite{Coulomb-Width-W} or slepton
 pair production at threshold\cite{Zerwas-threshold}. In our case
 width effects are of no importance since the characteristic
time of the Coulomb interaction, $1/\mstauo v^2$ typical of
velocities at freeze-out is  much smaller than the decay time,
$1/\Gamma_{\stauo}$, since in our example $\Gamma_{\stauo}/\mstauo
\simeq 2\; 10^{-5}$ and $v^2 \sim 0.2$. For smaller $\delta
m=\mstauo-\mneuto$, the width effect is even more negligible
whereas for larger $\delta m$, the $\stau_1 \stau_1$ channel would
be thermodynamically irrelevant. Therefore in our particular case
the resummation can be taken from the old Sommerfeld result.
Nonetheless, especially after thermal averaging, in our case this
type of QED correction is well approximated by the one-loop
approximation\footnote{ In a situation, which is not the case
here, where the $\stau^\pm \stau^\pm$ and $\stau^\pm \stau^\mp$
would contribute equally at tree-level, the Coulomb-Sommerfeld
correction would cancel after adding the two-channels. The
correction in  $\stau^\pm \stau^\mp$ would be attractive and given
by changing the sign of $X_{nr}$ in Eq.~\ref{resum-Sommerfeld}.The
non-perturbative effects of the Coulomb-Sommerfeld-like
corrections that might occur when coloured states are
involved\cite{BaerSommerfeld,Freitas-relic-qcd } need more care
because of the strong QCD coupling, bound states effects might be
relevant. The effect of the latter is even more important with
models with TeV and multi TeV dark matter almost degenerate with a
charged component\cite{Hisano-relic,Cirelli-Strumia-relic}. The
non perturbative effects on indirect detection in these models is
even more
dramatic\cite{Nojiri-gammaray-coulomb,Cirelli-Strumia-relic}.}.

Taking now all the effects and contributions in our specific
example we find an overall correction of $-8.6\%$ to the relic
density with a corrected value of $\Omega h^2=0.117$. As we can
see this value would not have been approached with a naive overall
rescaling of the effective couplings. Nonetheless, apart from some
$4\%$ correction, most of the effect seems to be explained in
terms of a proper usage of effective couplings and the Coulomb
effect. In fact in the total contribution, the Coulomb effect is
diluted and changes the results for the relic  density by about
only $1.5\%$.

\section{Annihilation of a mixed gaugino-higgsino LSP into vector bosons}
Having studied annihilation into fermions, annihilation into the
weak vector bosons is quite interesting. In the context of mSUGRA
this occurs for example in the so-called focus point region. In
order not to mix issues, we do not consider in this letter a
scenario where the LSP neutralino is either dominantly higgsino or
wino, therefore avoiding that $\neuto \chargop$ annihilation is of
relevance in  this case. We seek a scenario with a neutralino
where the largest component is still bino but where one has a
substantial higgsino and wino component. In our example one has
$\neuto= 0.819\tilde{b} -0.231 \tilde{w} -0.470\tilde{h}_1
-0.232\tilde{h}_2$ with $\mneuto=102.89,
\mchargo=125.13,\mchargt=215.27\;$GeV.
  All other masses
(outside the chargino-neutralino sector) are taken to be very
heavy at $1$TeV. This is also to avoid contamination from
annihilation into fermions. It would however be worth to study the
impact of the sfermions on the radiative corrections. In this case
we have not made the extra effort of searching for a set with a
relic density within the WMAP range. Here annihilation into $WW$
and $ZZ$ accounts for $80\%$, see Table~\ref{tab-mix1} with a few
other channels below $2\%$ each. These involve $\neuto \chargop$
into light quarks or $WZ$ which we take into account only at
tree-level.

\begin{table*}[h]
\begin{center}
\begin{tabular}{|c|c|c|c|c|c|}
\hline $\neuto\neuto\rightarrow W^+ W^-$ ($75\%$) &  Tree & & $A_{\tau \tau}$ & $\drbar$ & $MH$ \\
\hline a & 3.099 & & -27\% & -2\% & +44\% \\
\hline b & 5.961 & & -32\% & -7\% & +38\% \\
\hline
\hline $\neuto\neuto\rightarrow Z Z$ ($5\%$)&  Tree & & $A_{\tau \tau}$ & $\drbar$ & $MH$ \\
\hline a & 0.159 & & -22\% & +3\% & +50\% \\
\hline b & 0.787 & & -30\% & -6\% & +39\% \\
\hline
\hline $\Omega h^{2}$ & 0.053 & & 0.068 & 0.054 & 0.039\\
\hline $\frac{\delta \Omega h^{2}}{\Omega h^{2}}$ &  & & +28$\%$ & +2$\%$ & -26$\%$\\
\hline
\end{tabular}
\caption{{\em Mixed case: As in Table~\ref{tab-bino}. Note again
that the percentages in the first column next to the process refer
to the percentage weight, at tree-level, of that particular
channel.} \label{tab-mix1} }
\end{center}
\end{table*}

\begin{figure*}[htbp]
\begin{center}
\includegraphics[width=0.85\textwidth,height=0.6\textwidth]{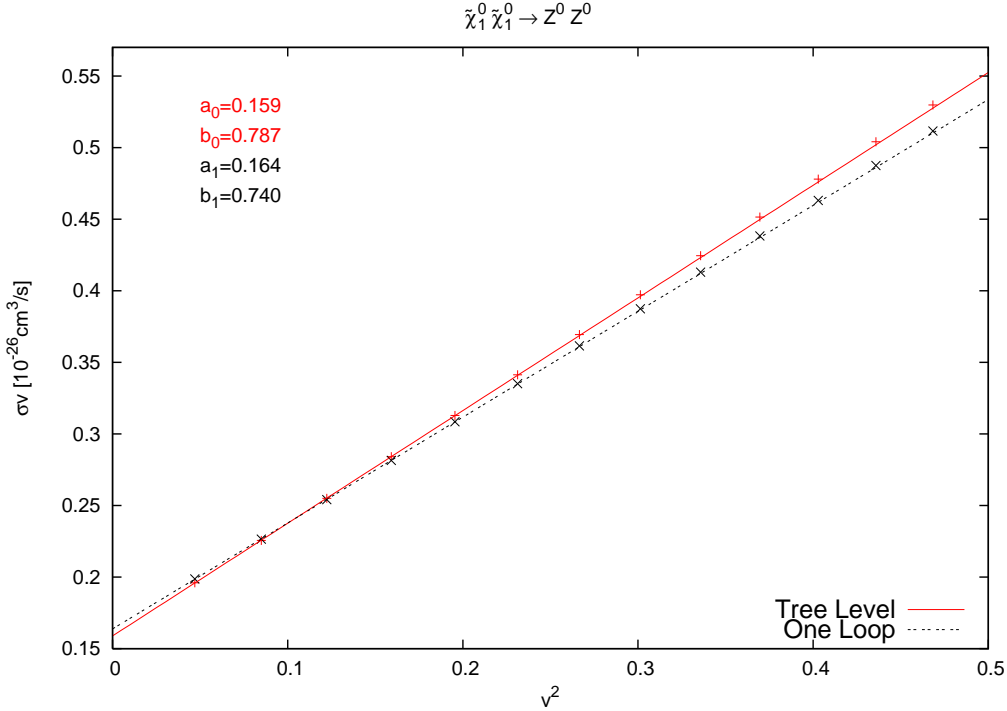}
\caption{\label{fig.mix-zz}{\em  $\neuto \neuto \ra Z Z$ as a
function of the square of the relative velocity both for
tree-level (solid line) and at one-loop (dotted line) in the
$\overline{DR}$ scheme. Fits to the $s$-wave ($a$), $p$-wave ($b$)
are indicated.}}
\end{center}
\end{figure*}

First of all, we see that the corrections which affect annihilation
into $ZZ$ and $WW$ are about the same (within $5\%$ in the 3 \tgbt
schemes). Moreover the correction to the $s$-wave and $p$-wave are
of the same order, see also Fig.~\ref{fig.mix-zz} for the $v$
dependence of the $ZZ$ cross section and the extraction of the
$s$- and $p$-wave coefficients. However this is not the most
important conclusion. The most important lesson is that there is a
very large \tgbt scheme dependence. In some other investigation
concerning Higgs decays, we had noticed, as was also pointed in
\cite{Freitas-Stockinger-tb} with a similar scheme, that the $MH$
scheme can lead to very large corrections. However in many
instances, like what we saw in the case of the bino annihilation
or co-annihilation, the $\delta \tgb$ is screened. Unfortunately
in this mixed neutralino scenario, the $\delta \tgb$ dependence
can be potentially enhanced by $1/(\mu^2-M_2^2)$ from the
renormalisation of $\delta \mu$ for example. This needs further
investigation. In this model, already at tree-level, we had
noticed that the cross sections were very sensitive to small
changes in the underlying parameters. Apart from the \tgbt scheme
dependence, the corrections in the $\drbar$ scheme look modest
especially for the dominant $s$- wave contribution. However one
should not forget that these small corrections are within the use
of $\alpha$ in the Thomson limit. Switching to a scale of order
$M_Z$, the corrections are large of order $20\%$. Therefore our
conclusion is that in such a scenario there are genuine large
corrections in all three schemes we have considered. This also
confirms the study of the chargino neutralino system at one-loop
in \cite{chargino-neutralino-rcvienna} which  showed that though
the corrections to the masses are modest, there can be a large (of
order $30\%$) change in the gaugino-higgsino component and hence a
large impact on cross sections.

\section{Annihilation of neutralinos to $b \bar b $ and a not too heavy pseudo-scalar Higgs}
We expose this last example for illustrative purposes. Indeed the
one-loop perturbative treatment of the Higgs coupling to a bottom
pair using the bottom pole mass, here we have taken all along
$m_b=4.7\;$GeV, is far from describing the bulk of the radiative
QCD corrections which as we know need to be resummed both for the
running effective $m_b$ purely from QCD and from the so-called
$\Delta m_b$ effects. The latter being more important for high
\tgbt. These effects are already taken in {\tt
micrOMEGAs}\cite{micromegas} for example. The purpose here
therefore is to see whether there are other possible effects,
though smaller, that are captured in a complete one-loop
calculation. Ideally one would like to subtract the known
universal one-loop QCD corrections from the full one-loop QCD
corrections that can occur for example from box diagrams, for
$v\ne 0$, outside the Higgs resonance. By the way because of the
Majorana nature of the LSP, at the smallest velocities the most
important Higgs resonance is the pseudo-scalar Higgs, $A$. In any
case for a precise calculation of the relic density, non resonant
contributions should be taken into account as thermal average is
to applied and would bring some smearing. Therefore, here we
concentrate on $\neuto\neuto\rightarrow b \bar b$ where the scalar
resonance is not negligible. Again like we argued for the previous
example we have to rely here also on some higgsino component. The
composition of the LSP is $\neuto= 0.896\tilde{b} -0.161 \tilde{w}
-0.378\tilde{h}_1 -0.166\tilde{h}_2$ with
$\mneuto=105.74,\mchargo=130.99,\mchargt=225.10\;$GeV. At
tree-level the system is (re)constructed from a set
${M_1,M_2,\mu}={110,150,-180}\;$GeV and $\tgb=5$. Compared to the
previous case where all other masses were around $1$TeV, to bring
out the effect of the $b$'s in the final state we lower first
$M_A$ such that $M_A=300\;$GeV. The masses of all sfermions are
around $200\;$GeV for the dominant $\tilde{f}_R$ and $250\;$GeV
for the dominant $\tilde{f}_L$. The mass of the gluino is also
lowered to be $400\;$GeV. \\

\noi At tree-level with $m_b=4.7\;$GeV, the dominant modes are
annihilation into $WW$ followed by $b \bar b$ (about $10\%$
smaller). $\tau \tau$ and $ZZ$ are about the same level but a
factor $10$ smaller than the $WW$ channel. We show, see
Table~\ref{tab-mix2}, the $WW$ and $ZZ$ channel in order to make a
comparison with the previous case of a mixed bino with a
substantial higgsino component. Here the \tgbt scheme dependence
has considerably reduced especially between the $A_{\tau \tau}$
and $\drbar$ scheme.  Note also that in the case of $\tau \tau$
channel there is also a discrepancy with the $MH$ scheme and the
other two schemes compared to the almost pure bino case. This
again is due to the larger contribution of the higgsino \tgbt
dependent part, naturally in the $A$-exchange not present in the
pure bino case but also in the $\tilde{\tau}$ exchange.

\begin{table}
\begin{center}
\begin{minipage}[t]{.4\linewidth}
\begin{tabular}{|c|c|c|c|c|c|}
\cline{1-1}
$\neuto\neuto\rightarrow$ & \multicolumn{5}{c}{ } \\
\hline $ W^+ W^-$ &  Tree & & $A_{\tau \tau}$ & $\drbar$ & $MH$ \\
\hline a & 0.904 & & -9\% & -3\% & +34\% \\
\hline b & 1.714 & & -10\% & -5\% & +30\% \\
\hline
\hline $ ZZ$  &  Tree & & $A_{\tau \tau}$ & $\drbar$ & $MH$ \\
\hline a & 0.061 & & +2\% & +5\% & +31\% \\
\hline b & 0.254 & & -6\% & -2\% & +24\% \\
\hline
\hline $ b \overline{b}$  &  Tree & & $A_{\tau \tau}$ & $\drbar$ & $MH$ \\
\hline a & 0.858 & & -27\% & -23\% & +5\% \\
\hline b & 1.032 & & -31\% & -27\% & -1\% \\
\hline
\hline $\tau^+ \tau^-$   &  Tree & & $A_{\tau \tau}$ & $\drbar$ & $MH$ \\
\hline a & 0.033 & & +3\% & +9\% & +52\% \\
\hline b & 0.631 & & +19\% & +18\% & +12\%\\
\hline
\end{tabular}
\end{minipage}
\hfill
\begin{minipage}[t]{.4\linewidth}
\begin{tabular}{|c|c|c|c|}
\hline $\neuto\neuto\rightarrow b \overline{b}$ & $A_{\tau \tau}$  & $\drbar$  & $MH$\\
\hline $\delta a/a$ EW & -1$\%$ & +3$\%$ & +31$\%$ \\
\hline $\delta a/a$ QCD & -26$\%$ & -26$\%$ & -26$\%$ \\
\hline
\hline $\delta b/b$ EW &  -1$\%$ & +3$\%$ & +29$\%$ \\
\hline $\delta b/b$ QCD & -30$\%$ & -30$\%$ & -30$\%$ \\
\hline
\end{tabular}
\end{minipage}
\caption{{\em Mixed case 2: As in Table~\ref{tab-mix1} for the
array on the left. The array on the right gives the relative
corrections to the $b \bar b$ channel for the QCD and EW
corrections. } \label{tab-mix2} }
\end{center}
\end{table}

\noi Let us look at what we obtain for the $b \bar b$ channel. The
electroweak corrections do show some \tgbt scheme dependence.
Compared to the electroweak corrections in the $\drbar$ scheme of
\tgbt, the QCD corrections are larger by an order of magnitude,
they amount to about $-30\%$ in both the $s$- and $p$-wave. If one
{\em assumes} that most of these corrections arise from the $A \ra
b \bar b$ vertex, then we know that there are large logs resulting
from the anomalous dimension of the pseudoscalar (and for that
matter the scalar). Its one loop part can be found in
\cite{Dress-qcd-aqq} and amounts to \beqn -\frac{4 \alpha_s}{\pi}
\left(\ln\frac{2\mneuto}{m_b}-3/4 \right). \nonumber \eeqn There
is also an important SUSY QCD correction termed $\Delta
m_b$\cite{delta_mb}, see section B.4.2 in the second paper
of\cite{micromegas}. In this scenario, adding these two
corrections amounts to about $-35\%$. Subtracting these from the
correction we calculate for the $s$-wave, leaves us with about
$+9\%$ QCD corrections. The QCD corrections from the anomalous
dimension and the $\Delta m_b$ effect were extracted for the known
effect to $A \ra b \bar b$. However, since at $v=0$, for the
s-wave cross section, the neutralino system constructs a
pseudoscalar because of its Majorana nature, the same corrections
should affect the $a$ coefficient even if the  contribution from
the pseudo-scalar Higgs is negligible.

\section{Conclusions}
We have performed the first electroweak corrections to some
important processes relevant for the relic density of neutralinos
in supersymmetry. This has become possible thanks to an automated
code for the calculation of loop corrections in the MSSM that will
allow to perform with the same tools and the same framework
(scheme dependence,..) analyses at one-loop at the collider and
for dark matter. Our findings suggest that in the case of a
dominantly bino neutralino, a large part of the correction can be
accounted for through an effective electromagnetic coupling at the
scale of the neutralino mass. Even so, complete one loop
corrections would be needed to match the foreseen precision of
PLANCK. The corrections to the relic density are not sensitive
much to the way \tgbt is renormalised. In the case of
co-annihilation of a bino and stau, the conclusion is similar but
one has to be wary of possible Coulomb-Sommerfeld corrections. For
a neutralino LSP which is strongly mixed, the corrections are
large and the \tgbt scheme dependence not negligible at all. More
investigation of such scenarii should be conducted. Some QCD (and
SUSY QCD) corrections affecting final states quarks in the case of
neutralino annihilation need that one goes beyond one-loop. Some
of these corrections have been identified and already implemented
in a code such as {\tt micrOMEGAs}. Apart from these corrections,
there remain some additional one-loop corrections that should be
taken into account. Before generalising these conclusions, more
work is needed. However, the tools exist. The next step is to
interface our code for the loop calculations with a dedicated
relic density calculator, avoiding double counting of some of the
one-loop corrections implemented as effective operators in the
relic density calculator. Work in this direction has already begun
based on {\tt micrOMEGAs} as the relic density calculator.

{\bf Acknowledgments} We would like to thank Sacha Pukhov for many
enlightening discussions especially concerning the future
implementation of the loop corrections into {\tt micrOMEGAs}.
David Temes has been invaluable in the first stages of the project
and gets all our thanks. We also owe much to our friends of the
Minami-Tateya group and the {\tt Grace-SUSY} code, in particular
we learned much from Masaaki Kuroda and were able to cross check
some one-loop results pertaining to decays of supersymmetric
particles. As usual, we thank Genevi\`eve B\'elanger for advice
and fruitful discussions. This work is supported in part by
GDRI-ACPP of the CNRS (France). The work of A.S. is supported by
grants of the Russian Federal Agency of Science NS-1685.2003.2 and
RFBR 04-02-17448. This work is also part of the French ANR
project, {\tt ToolsDMColl}.

\end{document}